%% file: paper.tex
\documentclass[prd,twocolumn,showpacs,groupedaddress,floatfix]{revtex4}

\usepackage{epsfig}
\usepackage{graphics}  
\usepackage{makeidx}
\usepackage{colordvi}
\usepackage{amsmath}
\usepackage{multirow}
\RequirePackage{xspace}
\usepackage{relsize}

\def\pep2{PEP-II\xspace}

\def\babar{\mbox{\slshape B\kern-0.1em{\smaller A}\kern-0.1em
    B\kern-0.1em{\smaller A\kern-0.2em R}}\xspace}

\def\invfb   {\ensuremath{\mbox{\,fb}^{-1}}\xspace}
\def\epem       {\ensuremath{e^+e^-}\xspace}

\def\CP                {\ensuremath{C\!P}\xspace}
\def\CPV                {\ensuremath{C\!PV}\xspace}
\def\CPT               {\ensuremath{C\!PT}\xspace} 

\def\FB                {{\scriptscriptstyle \rm FB}\xspace}

\def\qqbar {\ensuremath{q\overline q}\xspace}

\def\Y#1S{\ensuremath{\Upsilon{(#1S)}}\xspace}
\def\FourS {\Y4S}

\def\Bbar    {\kern 0.18em\overline{\kern -0.18em B}{}\xspace}

\def\BB      {\ensuremath{B\Bbar}\xspace} 

\def\Dz      {\ensuremath{D^0}\xspace}

\def\Kbar  {\kern 0.2em\overline{\kern -0.2em K}{}\xspace}

\def\Kz    {\ensuremath{K^0}\xspace}
\def\Kzb   {\ensuremath{\Kbar^0}\xspace}
\def\KzKzb {\ensuremath{\Kz \kern -0.2em - \kern -0.2em \Kzb}\xspace}
\def\KS    {\ensuremath{K^0_{\scriptscriptstyle S}}\xspace} 
\def\KL    {\ensuremath{K^0_{\scriptscriptstyle L}}\xspace} 
\def\KSKL {\ensuremath{\KS \kern -0.2em - \kern -0.2em \KL}\xspace}
\def\Kp    {\ensuremath{K^+}\xspace}
\def\Km    {\ensuremath{K^-}\xspace}
\def\Kpm   {\ensuremath{K^\pm}\xspace}

\def\pip  {\ensuremath{\pi^+}\xspace}
\def\pim  {\ensuremath{\pi^-}\xspace}
\def\pipm  {\ensuremath{\pi^\pm}\xspace}

\def\epm  {\ensuremath{e^\pm}\xspace}
\def\ppm  {\ensuremath{p^\pm}\xspace}

\newcommand{\gevc}{\ensuremath{{\mathrm{\,Ge\kern -0.1em V\!/}c}}\xspace}
\newcommand{\mevc}{\ensuremath{{\mathrm{\,Me\kern -0.1em V\!/}c}}\xspace}
\newcommand{\gevcc}{\ensuremath{{\mathrm{\,Ge\kern -0.1em V\!/}c^2}}\xspace}
\newcommand{\mevcc}{\ensuremath{{\mathrm{\,Me\kern -0.1em V\!/}c^2}}\xspace}

\newcommand{\stat}{\ensuremath{\mathrm{(stat)}}\xspace}
\newcommand{\syst}{\ensuremath{\mathrm{(syst)}}\xspace}

\newcommand{\Dtokspi}{\ensuremath{D^{\pm}\to\KS\pipm}\xspace}
\newcommand{\Dtoksk}{\ensuremath{D^{\pm}\to\KS\Kpm}\xspace}
\newcommand{\Dstoksk}{\ensuremath{D_s^{\pm}\to\KS\Kpm}\xspace}
\newcommand{\Dstokspi}{\ensuremath{D_s^{\pm}\to\KS\pipm}\xspace}

\newcommand{\Dps}{\ensuremath{D_{(s)}}\xspace}
\newcommand{\Dpstoksk}{\ensuremath{\Dps^\pm \to \KS \Kpm}\xspace}

\begin{document}  

\begin{flushleft}
\babar-PUB-12/029\\
SLAC-PUB-15318\\
\end{flushleft}

\title{
{\large  \boldmath
Search for \CP Violation in the Decays \Dtoksk, \Dstoksk, and \Dstokspi}
}

\input{authors_oct2012_bad2480}

\begin{abstract} 
We report a search for \CP violation in the decay modes \Dtoksk, \Dstoksk, and \Dstokspi
using a data set corresponding to an integrated luminosity of $469\,\invfb$ 
collected with the \babar detector at the \pep2 asymmetric energy \epem storage rings.
The decay rate \CP asymmetries, $A_{\CP}$, are determined to be
$(+0.13 \pm 0.36 \stat \pm 0.25 \syst)\%$, 
$(-0.05 \pm 0.23 \stat \pm 0.24 \syst)\%$, and 
$(+0.6 \pm 2.0 \stat \pm 0.3 \syst)\%$, respectively.
These measurements are consistent with zero, 
and also with the standard model prediction ($(-0.332 \pm 0.006)\%$
for the \Dtoksk and \Dstoksk modes, and $(+0.332 \pm 0.006)\%$ for the \Dstokspi mode).
They are the most precise determinations to date.
\end{abstract}


\vspace{-0.2cm}
  
\maketitle    

\section{Introduction}

The search for \CP violation (\CPV) in charm decays provides 
a sensitive probe of physics beyond the Standard Model (SM). 
Owing to its suppression within the SM, 
a significant observation of direct \CPV in charm decays 
would indicate the possible presence of new physics (NP) effects in the decay processes.   
In a previous article~\cite{delAmoSanchez:2011zza}, 
we reported a precise measurement of the \CP asymmetry in the \Dtokspi mode, 
where the measured asymmetry was found to be consistent with the 
value expected from indirect \CPV in the \Kz system. 

The LHCb and CDF Collaborations have recently reported evidence for \CPV in
charm decays by measuring the difference of \CP asymmetries in the
$\Dz \to \Kp \Km$ and $\Dz \to \pip \pim$ channels~\cite{Aaij:2011in,Collaboration:2012qw},
which is mainly sensitive to direct \CPV. 
The size of the world average direct \CP asymmetry difference,
$(-6.56\pm 1.54)\times 10^{-3}$~\cite{Amhis:2012bh}, suggests
either a significant enhancement of SM penguin amplitudes 
or of NP amplitudes (or both) in charm decays~\cite{Isidori:2011qw_etal}.
Improved measurements of the \CP asymmetries in the individual two-body modes, 
along with measurements in other channels, are needed to 
determine the nature of the contributing amplitudes.

We present herein measurements of the decay rate \CP asymmetry, $A_{\CP}$,
defined as
\begin{equation}
A_{\CP}=\frac{\Gamma(D^+_{(s)}\to f)-\Gamma(D_{(s)}^-\to \overline{f})}
{\Gamma(D^+_{(s)}\to f)+\Gamma(D^-_{(s)}\to \overline{f})},
\end{equation}
in the decay modes \Dtoksk, \Dstoksk, and \Dstokspi.
Previous measurements of $A_{\CP}$ in these channels have been reported by 
the CLEO-c~\cite{:2007zt} and Belle Collaborations~\cite{Ko:2010ng}.
As for the $A_{\CP}$ measurement in \Dtokspi,
we expect an $A_{\CP}$ asymmetry of $(\pm0.332\pm 0.006)\%$~\cite{Nakamura:2010zzi}
resulting from \CPV in $\Kz-\Kzb$ mixing~\cite{Lipkin:1999qz}.
The sign of the $\Kz$-induced asymmetry is positive (negative) if a \Kz (\Kzb)
is present in the corresponding tree-level Feynman diagram.
Because it is identified by its $\pi^+\pi^-$ decay, the intermediate state 
is a coherent mix of \KS and \KL amplitudes. It has been shown in Ref.~\cite{Azimov:1981ss} that the \KSKL
interference term gives rise to a measured \CP asymmetry that depends on the range in proper 
time over which the decay rates are integrated, and on the efficiency for the reconstruction of the 
intermediate state as a function of its proper flight time.

For this analysis we employ a technique similar to that used 
for our measurement of \CPV in the \Dtokspi mode~\cite{delAmoSanchez:2011zza}.
As a result, reference to our previous publication
is given for the description of some of the analysis details.

\section{The \babar detector and event selection}

The data used for these measurements were recorded at or near the
$\Y4S$ resonance by the \babar detector at the \pep2 storage rings,
and correspond to an integrated luminosity of $469\,\invfb$.
Charged particles are detected, and their momenta measured, by a combination of a silicon
vertex tracker, consisting of 5 layers of double-sided
detectors, and a 40-layer central drift chamber, both operating
in a 1.5 T axial magnetic field. 
Charged-particle identification is
provided by specific ionization energy loss measurements in the tracking system, and by the
measured Cherenkov angle from an internally reflecting ring-imaging
Cherenkov detector covering the central region of the detector. 
Electrons are detected by a CsI(Tl) electromagnetic calorimeter.
The \babar detector, and the coordinate system used throughout, 
are described in detail in Refs.~\cite{Aubert:2001tu,Menges:2006xk}.
We validate the analysis procedure using 
Monte Carlo (MC) simulation based on Geant4~\cite{Agostinelli:2002hh}.
The MC samples include $\epem\to\qqbar\;(q=u,d,s,c)$ events, simulated with 
JETSET~\cite{Sjostrand:2006za} and \BB decays simulated with the EvtGen 
generator~\cite{Lange:2001uf}.
To avoid potential bias in the measurements
we finalize the event selection for each channel,
as well as the procedures for efficiency correction, fitting, 
and the determination of the systematic uncertainties and possible biases in the 
measurements, prior to extracting the value of $A_{\CP}$ from the data.

Signal candidates are reconstructed by combining a $\KS$ 
candidate, reconstructed in the decay mode $\KS\to\pi^+\pi^-$, 
with a charged pion or kaon candidate.
A \KS candidate is reconstructed from two oppositely charged
tracks with an invariant mass within 
a $\pm 10\,\mevcc$ interval centered on the nominal \KS mass~\cite{Nakamura:2010zzi},
which is approximately $\pm 2.5\,\sigma$ in the measured
\KS mass resolution. The $\chi^2$-probability of 
the $\pi^+\pi^-$ vertex fit must be greater than $0.1\,\%$.
Motivated by MC studies, we require the measured flight length
of the \KS candidate to be at least three times greater than its uncertainty,
to reduce combinatorial background.
A reconstructed charged-particle track that has $p_T\ge 400\,\mevc$ 
is selected as a pion or kaon candidate, where $p_T$ is the magnitude  
of the momentum in the plane perpendicular to the $z$ axis (transverse plane).
In our measurement, we require that a pion candidate not be identified 
as a kaon, a proton, or an electron, and that a kaon candidate 
be identified as a kaon, and not as a pion, a proton, or an electron.
Identification efficiencies and misidentification rates for
electron, pions, kaons, and protons with 2 \gevc momentum in the laboratory frame
are reported in Table \ref{tab:pideff}.
\begin{table}[tb]
\caption{Identification efficiencies and misidentification rates for
electron, pions, kaons, and protons with 2 \gevc momentum in the laboratory frame.
The values for kaons on the third row refers to the identification criterion 
used to reject kaons from the pion sample, 
while the values on the fourth row to the criterion used in the kaon selection.}
\begin{center}
\begin{tabular}{|l|c|c|c|}\hline
\multirow{2}{*}{Particle}  & \multirow{2}{*}{Efficiency[\%]} & 
\multicolumn{2}{|c|}{Misid. rate [\%]}\\
\cline{3-4}
& & \pipm & \Kpm \\
\hline
\epm          & 91 & 0.04 & $<0.2$ \\
\pipm          & 88 &  $-$  &1 \\
\Kpm (applied to \pipm) & 91 & 1 & $-$ \\
\Kpm (applied to \Kpm) & 99 & 8 & $-$ \\
\ppm          & 80 & 0.2 & 0.2  \\
\hline
\end{tabular}
\label{tab:pideff}
\end{center}
\end{table}
The criteria used to select pion or kaon candidates are very effective in reducing
the charge asymmetry from track reconstruction and identification, as inferred from 
studying the data control samples described below.
A vertex fit to the whole decay chain, constraining the $\Dps^\pm$ production vertex to 
be within the \epem interaction region, is then performed~\cite{Hulsbergen:2005pu}.
We retain only $\Dps^\pm$ candidates having a $\chi^2$-probability for this fit
greater than 0.1\%, and an invariant mass $m(\KS h), h = \pi, K, $ within a $\pm65\mevcc$ 
interval centered on the nominal $\Dps^\pm$ mass~\cite{Nakamura:2010zzi},
which is approximately equivalent to $\pm 8\,\sigma$ in the measured
$\Dps^\pm$ mass resolution.

We require further that the magnitude of the $D^\pm_{s}$ candidate momentum in the 
$\epem$ center-of-mass (CM) system, $p^*$, 
be between 2.6 and 5.0 \gevc, in order to suppress combinatorial background from \BB events.
For the \Dtoksk mode, the MC simulated sample shows that
retaining candidates with $p^*$ between 2.0 and 5.0 \gevc
allows signal candidates from $B$-meson decays,
without introducing an excessive amount of combinatorial background.
Assuming that \CPT is conserved, there is no contribution to $A_{\CP}$ from \CP violation 
in $B$ meson decays from Standard Model processes.
Additional background rejection is obtained by requiring that the impact parameter 
of the $\Dps^\pm$ candidate with respect to the beam-spot~\cite{Aubert:2001tu}, 
projected onto the transverse plane,
be less than 0.3 cm, and that the $\Dps^\pm$ proper decay time, $t_{xy}$, 
be between $-15$ and $35$ ps. 
The decay time is measured using $L_{xy}$, defined as the distance of 
the $\Dps^\pm$ decay vertex from the beam-spot
projected onto the transverse plane.

In order to further optimize the sensitivity of the $A_{\CP}$ measurements, 
we construct a multivariate algorithm, based on seven 
discriminating variables for each $\Dps^\pm$ candidate:
$t_{xy}$, $L_{xy}$,
$p^*$, the momentum magnitude and component in the transverse plane
for the \KS candidate, and also for the pion or kaon candidate.
For the \Dtoksk and \Dstoksk modes the multivariate algorithm with the best performance 
is a Boosted Decision Tree~\cite{Speckmayer:2010zz},
while for the \Dstokspi mode the best algorithm is a Projective Likelihood method~\cite{Speckmayer:2010zz}.
The final selection criteria, based on the outputs of the multivariate selectors,
are optimized using truth-matched signal and 
background candidates from the MC sample.
For the optimization, we maximize the $S/\sqrt{S+B}$ ratio, where 
$S$ and $B$ are the numbers of signal and background candidates 
with invariant mass within $\pm 30 \mevcc$ of the nominal $\Dps^\pm$ mass,
which is approximately $\pm 3\,\sigma$ in the measured mass resolution. 

\section{Signal yield and asymmetry extraction}

\begin{figure*}[tb]
\begin{center}
\begin{tabular}{ccc}
\includegraphics[width=0.33\textwidth]{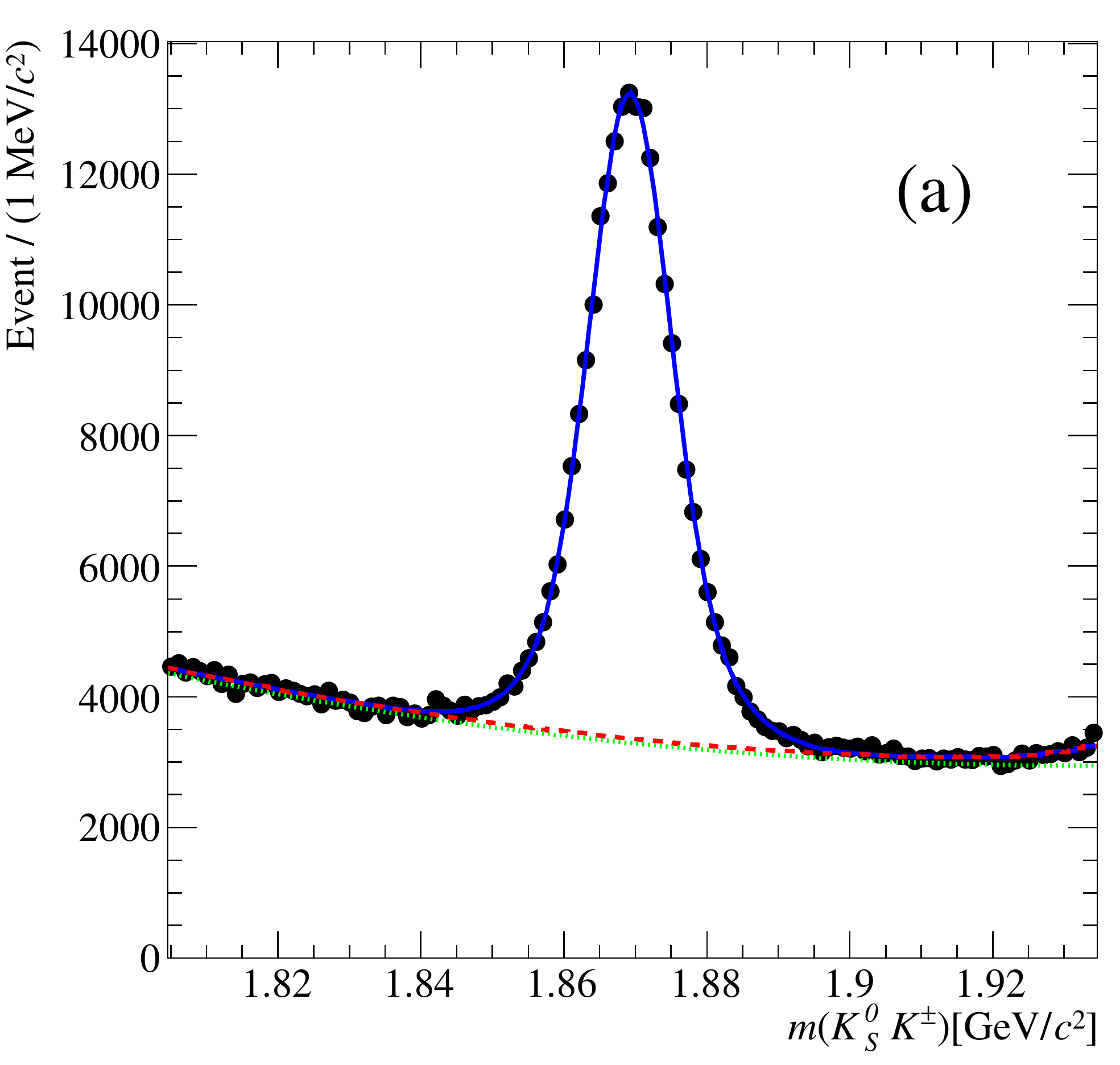} &
\includegraphics[width=0.33\textwidth]{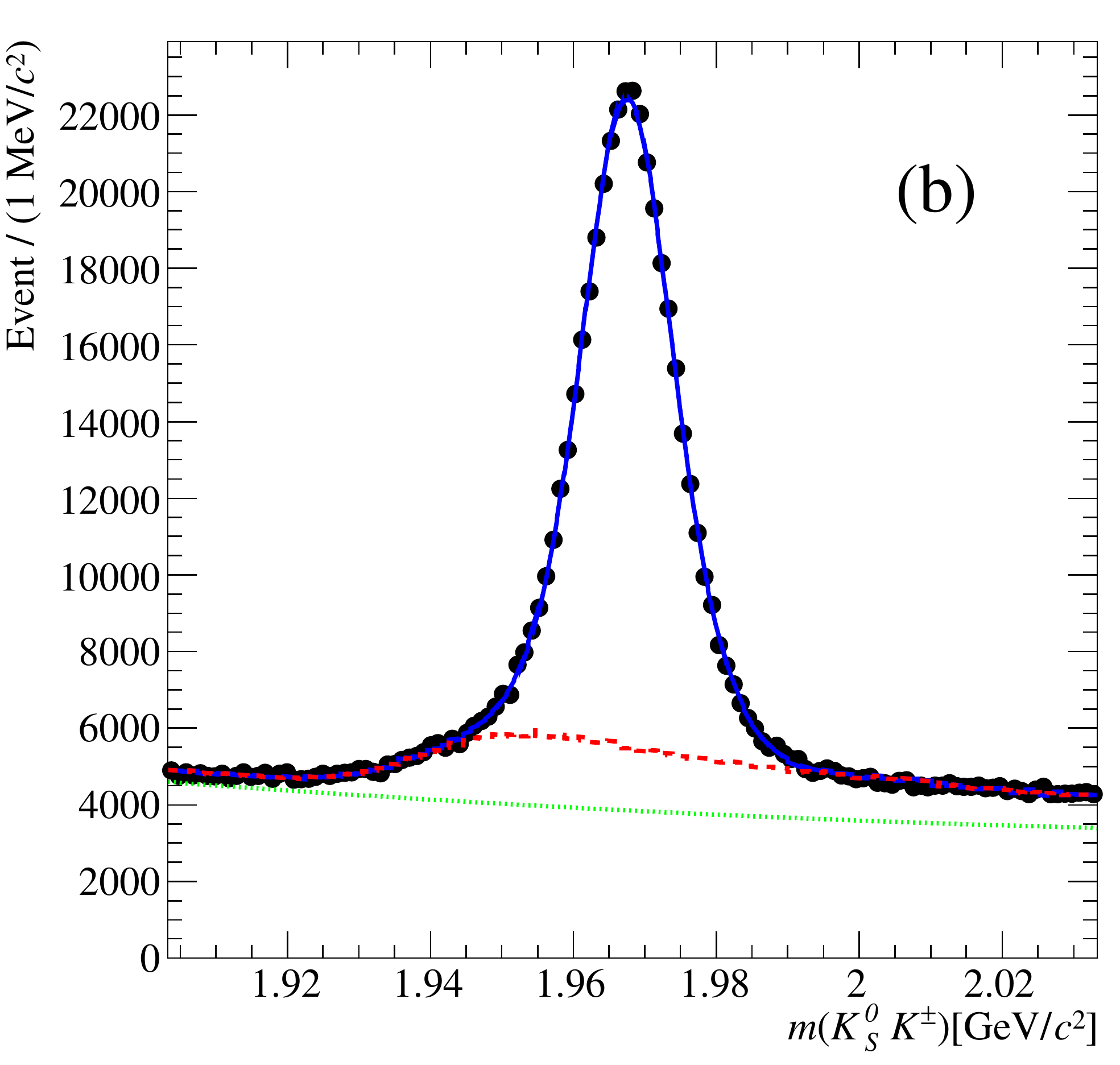} &
\includegraphics[width=0.33\textwidth]{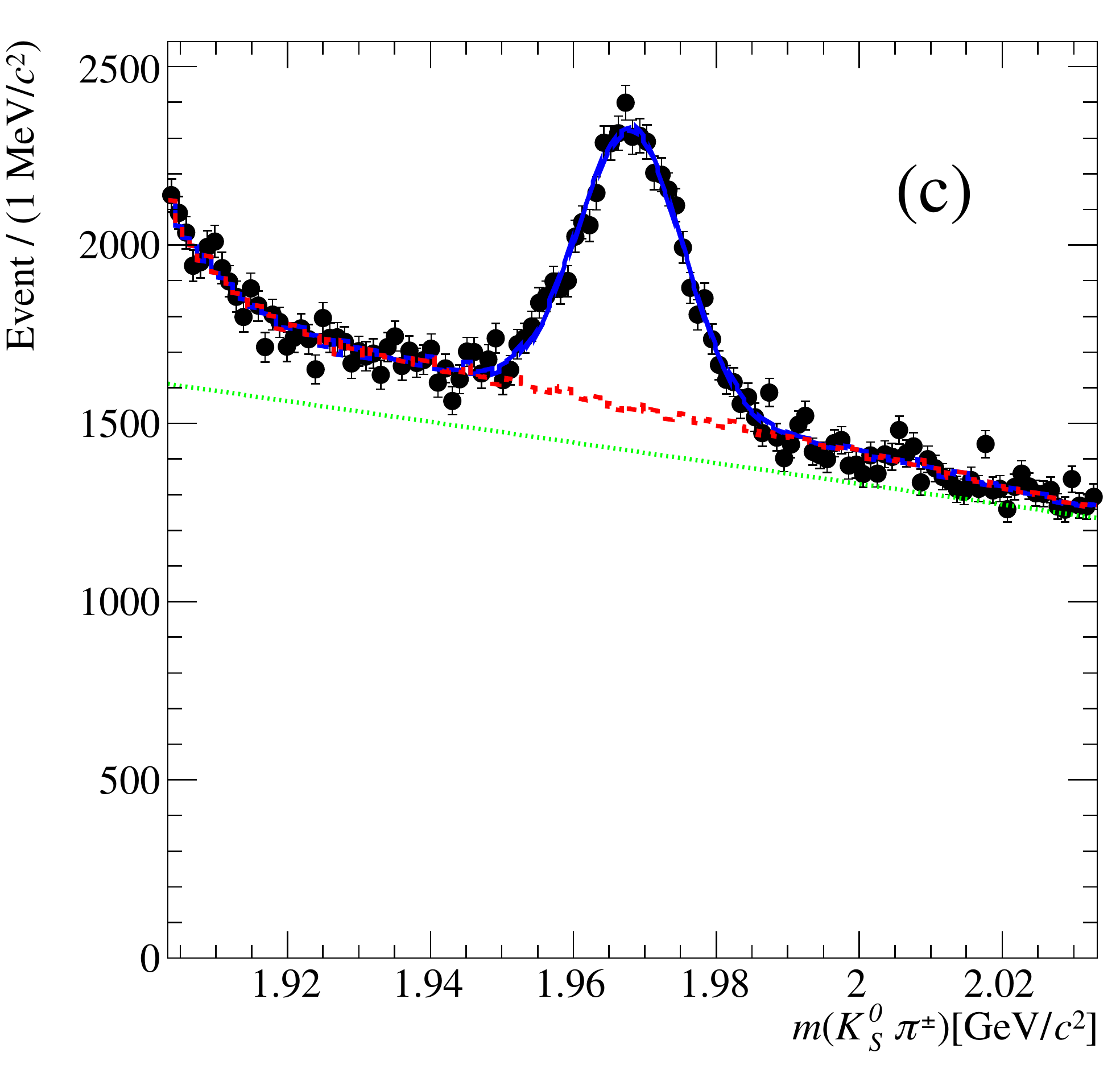} \\
\end{tabular}
\vspace{-0.3cm}
\caption{
Invariant mass distribution for (a) \Dtoksk, (b) \Dstoksk, and (c) \Dstokspi 
candidates (points with error bars).
The solid curve shows the result of the fit to the data.
The dashed curve represents the sum of all background contributions,
while the dotted curve indicates combinatorial background only.}
\label{fig1}
\vspace{-0.7cm}
\end{center}
\end{figure*}

For each mode the signal yield is extracted using 
a binned maximum likelihood (ML) fit to the distribution 
of the invariant mass $m(\KS h)$ for the
selected $\Dps^\pm$ candidates.
The total probability density function (PDF) 
is the sum of signal and background components. 
The signal PDF is modeled as a sum of two Gaussian functions 
for the \Dpstoksk modes, and as a single Gaussian function for the \Dstokspi mode.
The background PDF is taken as the sum of two components: a distribution describing 
the invariant mass of mis-reconstructed charm meson decays, 
and a combinatorial background modeling the mass distribution from other sources.
For the \Dtoksk (\Dstokspi) mode the charm background is mainly from the tail of the invariant mass 
distribution for \Dstoksk (\Dtokspi) candidates.
For the \Dstoksk mode, the mis-reconstructed charm background originates mainly 
from \Dtokspi decays for which the \pipm is misidentified as a \Kpm.
Assigning the wrong mass to the pion shifts the reconstructed invariant mass, and the resulting
distribution is a broad peak with mean value close to the $D_{s}^\pm$ mass.
For each mode, the invariant mass distribution due to charm background is modeled 
using a histogram PDF obtained from a MC sample of simulated charm background decays.
The combinatorial background is described by a first(second)-order polynomial for the \Dstokspi mode 
(\Dtoksk and \Dstoksk modes).
The fits to the $m(\KS h)$ distributions yield
$(159.4 \pm 0.8) \times 10^3$  \Dtoksk decays,
$(288.2 \pm 1.1) \times 10^3$  \Dstoksk decays, and
$(14.33 \pm 0.31) \times 10^3$ \Dstokspi decays.
The data and the fit results are shown in Fig.~\ref{fig1}.
All of the PDF parameters are extracted from fits to the data.

For each channel, we determine $A_{\CP}$ by measuring the signal yield asymmetry $A$ defined as:
\begin{equation}
A=\frac{N_{\Dps^+}-N_{\Dps^-}}{N_{\Dps^+}+N_{\Dps^-}},
\end{equation}
where $N_{\Dps^+}$($N_{\Dps^-}$) is the number of $\Dps^+$($\Dps^-$) decays determined from the fit
to the invariant mass distribution.
The asymmetry $A$ contains two contributions in addition to $A_{\CP}$, namely
the forward-backward (FB) asymmetry ($A_{\FB}$), and a detector-induced component.
We measure $A_{\FB}$ together with $A_{\CP}$ using the selected dataset, 
while we correct the data for the detector-induced component
using coefficients derived from a control sample.

\section{Correction of detector-related asymmetries}

\begin{figure}[tb]
\begin{center}
\includegraphics[width=0.47\textwidth]{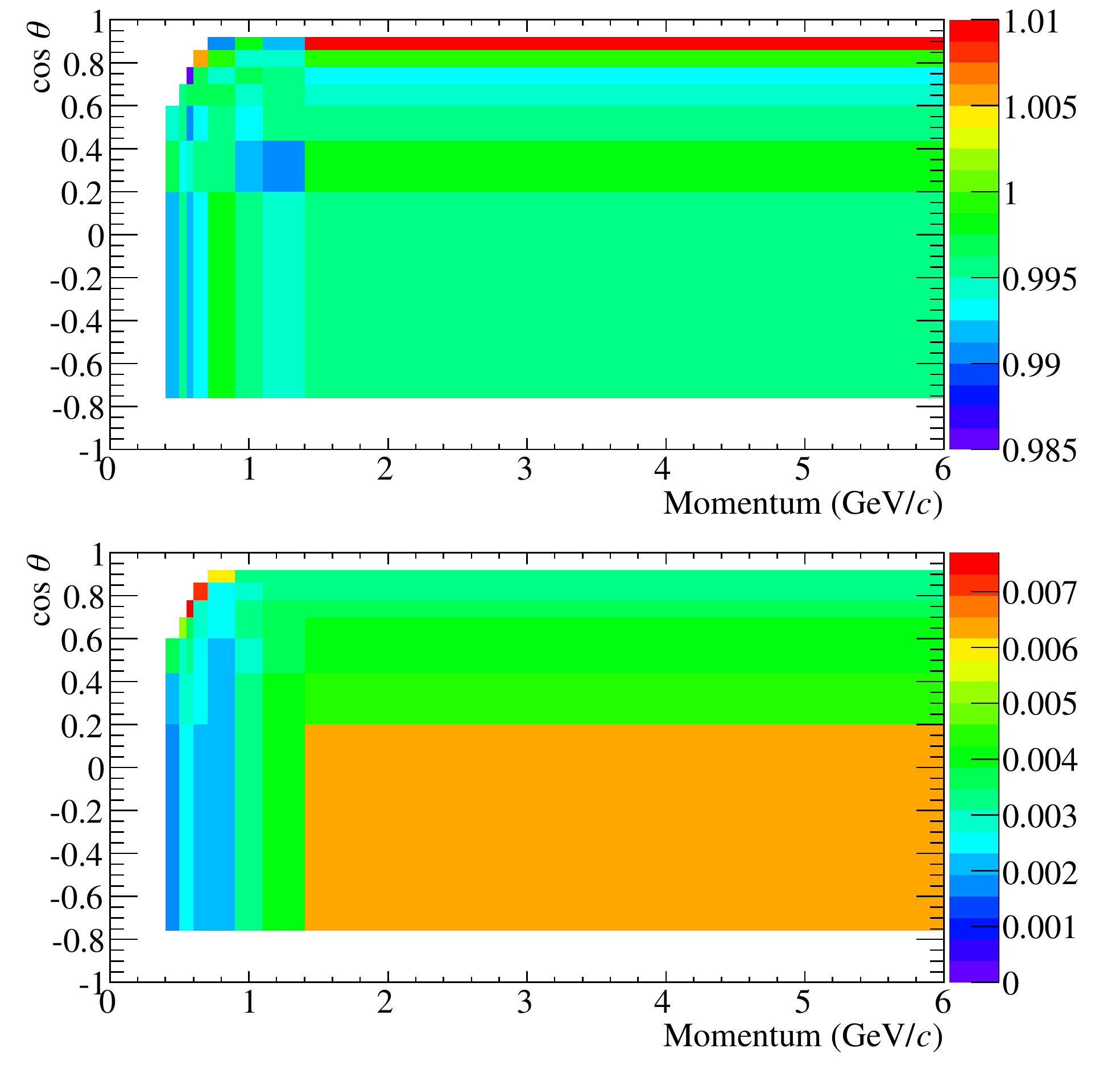}
\vspace{-0.3cm}
\caption{(top) The ratio between the detection efficiency
for $\pi^+$ and $\pi^-$, and (bottom) the
corresponding statistical errors. 
The values are computed using the numbers of $\pi^+$ and $\pi^-$ 
tracks in the selected control sample.}
\label{fig4}
\vspace{-0.7cm}
\end{center}
\end{figure}

\begin{figure}[tb]
\begin{center}
\includegraphics[width=0.47\textwidth]{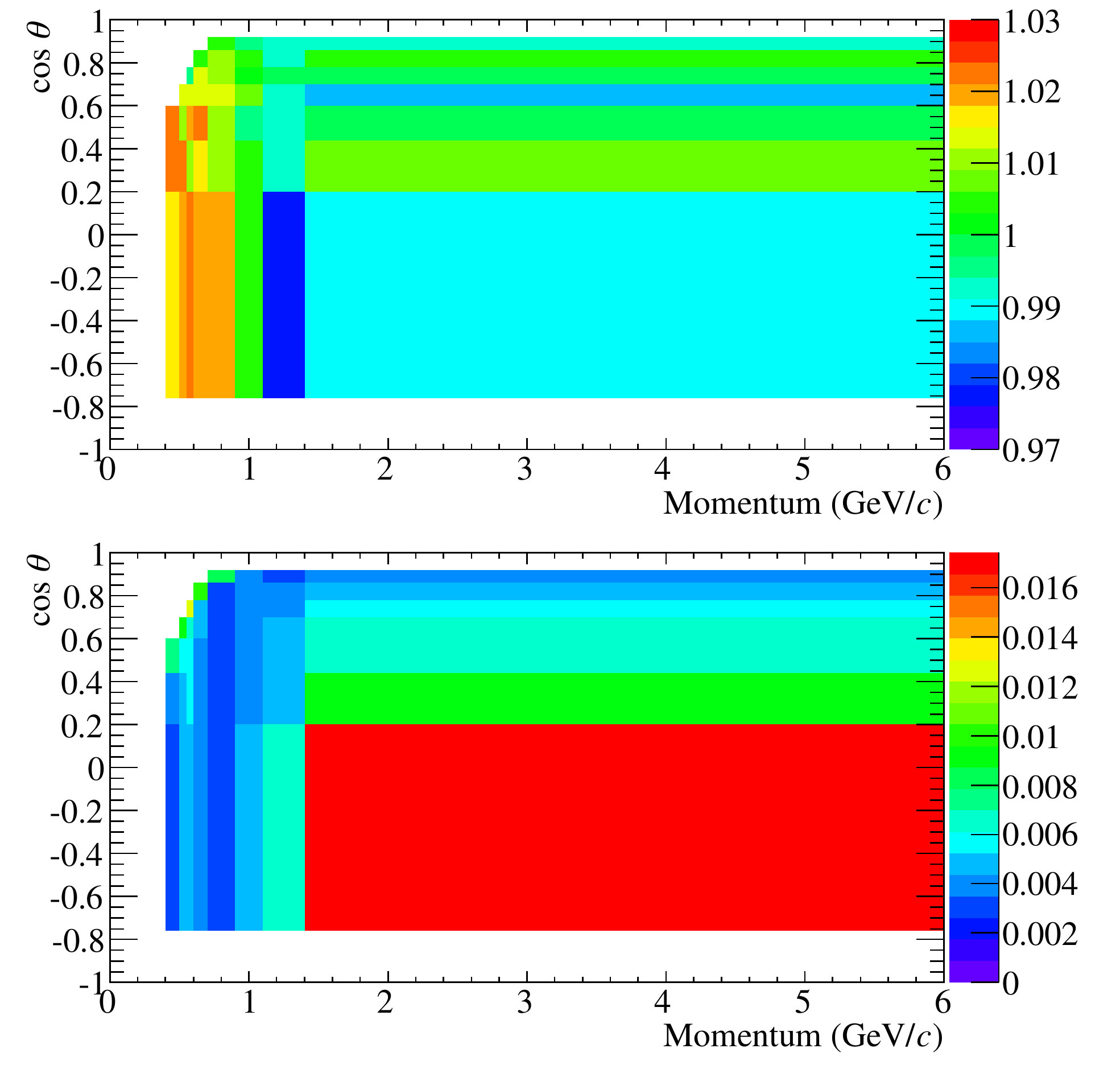}
\vspace{-0.3cm}
\caption{(top) The ratio between the detection efficiency
for $K^+$ and $K^-$, and (bottom) the 
corresponding statistical errors. 
The values are computed using the numbers of $K^+$ and $K^-$ 
tracks in the selected control sample.}
\label{fig5}
\vspace{-0.7cm}
\end{center}
\end{figure}

We use a data-driven method, described in detail in Ref.~\cite{delAmoSanchez:2011zza}, 
to determine the charge asymmetry in track reconstruction as a function of the magnitude of 
the track momentum and its polar angle in the laboratory frame. 
The method exploits the fact that $\FourS \to \BB$ 
events provide a sample evenly populated with positive and negative tracks, 
free of any physics-induced asymmetries.
The off-resonance momentum distribution is subtracted from the on-resonance one, 
to remove any contribution from continuum, for which there is a FB asymmetry in the CM frame.  
This sample is used to compute the detector-related asymmetries in the reconstruction of charged-particle tracks.
Starting from a sample of $50.6\,\invfb$ of data collected at the \FourS resonance and 
an off-resonance data sample of $44.8\,\invfb$, we obtain a large sample of charged-particle tracks,
and apply the same charged pion or kaon track selection criteria 
used in the reconstruction of the \Dpstoksk and \Dstokspi modes.
Then, after subtracting the off-resonance contribution from the on-resonance sample,
we obtain a sample of more than 120 million pion candidates, and 40 million kaon candidates, 
originating from \FourS decays.
We use the full off-resonance sample and an equivalent luminosity for
the on-resonance sample, because, due to the subtraction procedure, including additional
data in the on-resonance sample does not improve the statistical error 
on the correction ratios mentioned below.
These candidates are then used to compute the efficiency ratios 
for positive and negative pions and kaons.
The ratio values and their statistical errors for pions and kaons are shown 
in Fig.~\ref{fig4} and Fig.~\ref{fig5}, respectively.
For the $\Dps^{-} \to \KS \Km$ ($D_s^{-} \to \KS \pim$) modes,
the $\Dps^{-}$ ($D_s^{-}$) yields, in intervals of kaon (pion) momentum and
cosine of its polar angle, $\cos\theta$, are weighted with the kaon (pion) efficiency ratios to
correct for the detection efficiency differences between \Kp and \Km (\pip and \pim).
Momentum and cosine of its polar angle intervals are not uniform in order to have
similar statistics, and therefore similar correction uncertainty, in each interval. 
Interval sizes vary from (0.05 \gevc, 0.06) to (4.4 \gevc, 0.96),
where the first number is the momentum interval, and the second its cosine of polar angle interval.
The largest correction is approximately 1\% for pions and 2\% for kaons.  
After correcting the data for the detector-induced component only $A_{\FB}$ and $A_{\CP}$
contribute to the measured asymmetry $A$. 

\section{Extraction of $A_{\CP}$ and $A_{\FB}$}

Neglecting higher-order terms that contain $A_{\CP}$ and $A_{\FB}$,
the resulting asymmetry can be expressed simply as the sum of the two.
Given that $A_{\FB}$ is an odd function of $\cos\theta^*_D$, 
where $\theta^*_D$ is the polar angle of the $\Dps^\pm$ 
candidate momentum in the CM frame,
$A_{\CP}$ and $A_{\FB}$ can be written as a function of $|\cos\theta^*_D|$ as follows:
\begin{align}
  A_{\CP}(|\cos\theta^*_D|) &= \frac{A(+|\cos\theta^*_D|) + A(-|\cos\theta^*_D|)}{2}\\
  \intertext{and}
  A_{\FB}(|\cos\theta^*_D|) &= \frac{A(+|\cos\theta^*_D|) - A(-|\cos\theta^*_D|)}{2}, 
\label{eq:AcpAfb_intro}
\end{align}
where $A(+|\cos\theta^*_D|)$ ($A(-|\cos\theta^*_D|)$) is the measured asymmetry 
for the $\Dps^\pm$ candidates in a positive (negative) $\cos\theta^*_D$ interval.

A simultaneous ML fit to the $\Dps^+$ and $\Dps^-$ invariant mass distributions
is carried out to extract the signal yield asymmetry in each of
ten equally spaced $\cos\theta^*_D$ intervals, starting with interval 1 at $[-1.0,-0.8]$. 
The PDF model that describes the distribution in each sub-sample
is the same as that used in the fit to the full sample,
but the following parameters are allowed to float separately
in each sub-sample (referred to as split parameters):
the yields for signal, charm background and combinatorial candidates; 
the asymmetries for signal and combinatorial candidates;
the width, and the fraction of the Gaussian function with the larger contribution to the signal PDF;
and the first-order coefficient of the polynomial that models the combinatorial background.
For the \Dtoksk mode the yields for the charm background candidates in intervals 1, 2, and 3
were fixed to 0 to obtain a fully convergent fit.
Since interval 10 contains the smallest number of candidates,
we use a single Gaussian function to model the signal PDF for the \Dpstoksk modes.
For the \CP asymmetry of charm background candidates 
we use the same floating parameters as for the signal
candidates, because the largest source of \CP asymmetry for both samples is due to \CPV in
\KzKzb mixing. For the \Dstokspi mode, where the primary charm background channel, \Dtokspi, 
has the same magnitude but opposite-sign asymmetry due to \KzKzb mixing, we use a separate
parameter for the asymmetry of the charm background candidates. 
To achieve a more stable fit, 
if the fit results for a split parameter are statistically compatible 
between two or more sub-samples, 
the parameter is forced to have the same floating value among those sub-samples only.
For the \Dstokspi mode the width of the first Gaussian function for the signal PDF is set to the same
floating value in intervals 1, 2, 3, and 4.
The first-order coefficient of the polynomial describing the combinatorial background
is set to the same floating value in intervals 4 to 8 (\Dtoksk), in intervals 4 to 8 (\Dstoksk), and
in intervals 2 to 7 (\Dstokspi).
The final fit contains 70, 80, and 64 free parameters
for the \Dtoksk, \Dstoksk, and \Dstokspi modes, respectively.

\begin{figure*}[tb] \begin{center}
\begin{tabular}{ccc}
\includegraphics[width=0.33\textwidth,clip=true]{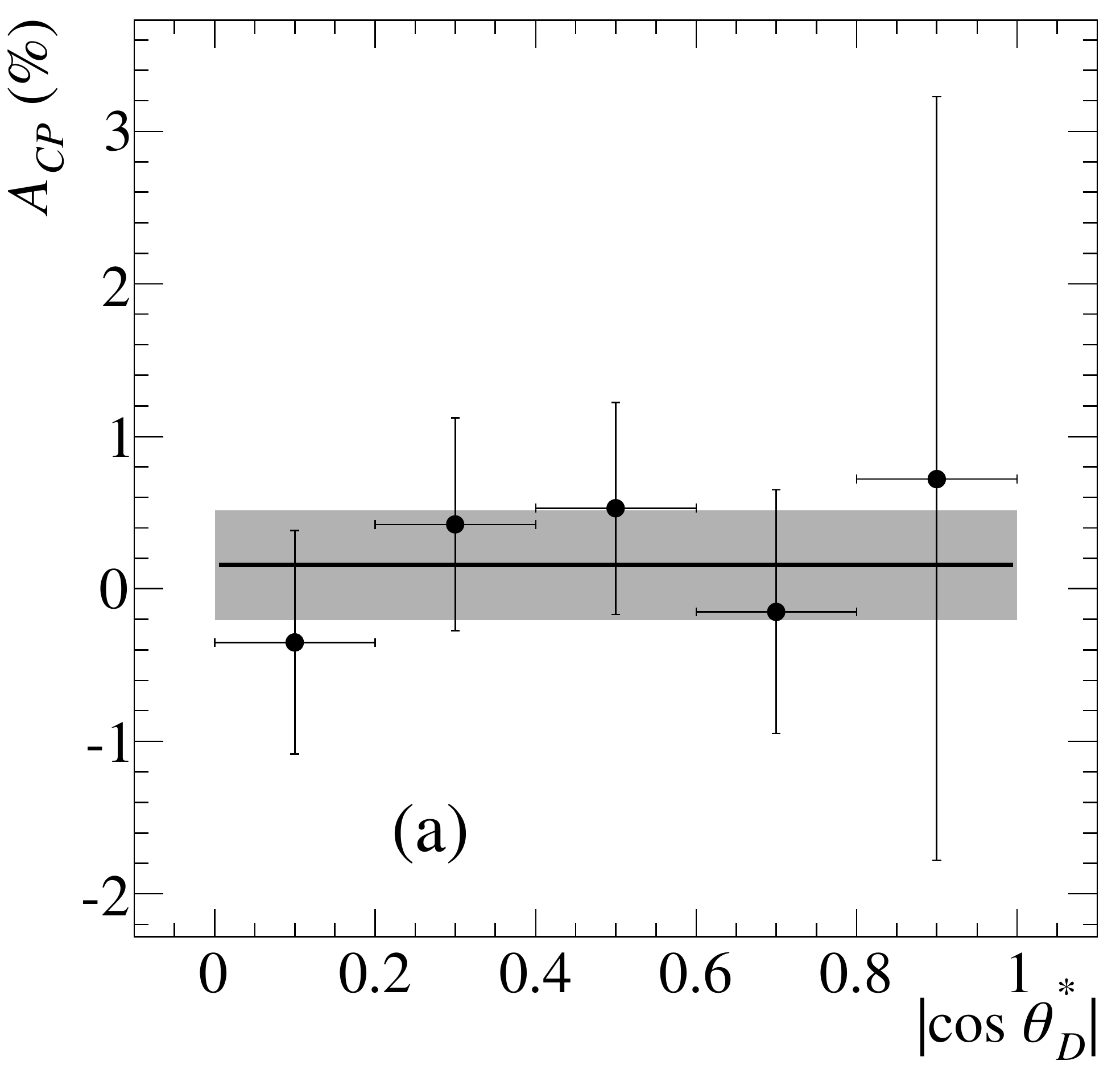} &
\includegraphics[width=0.33\textwidth,clip=true]{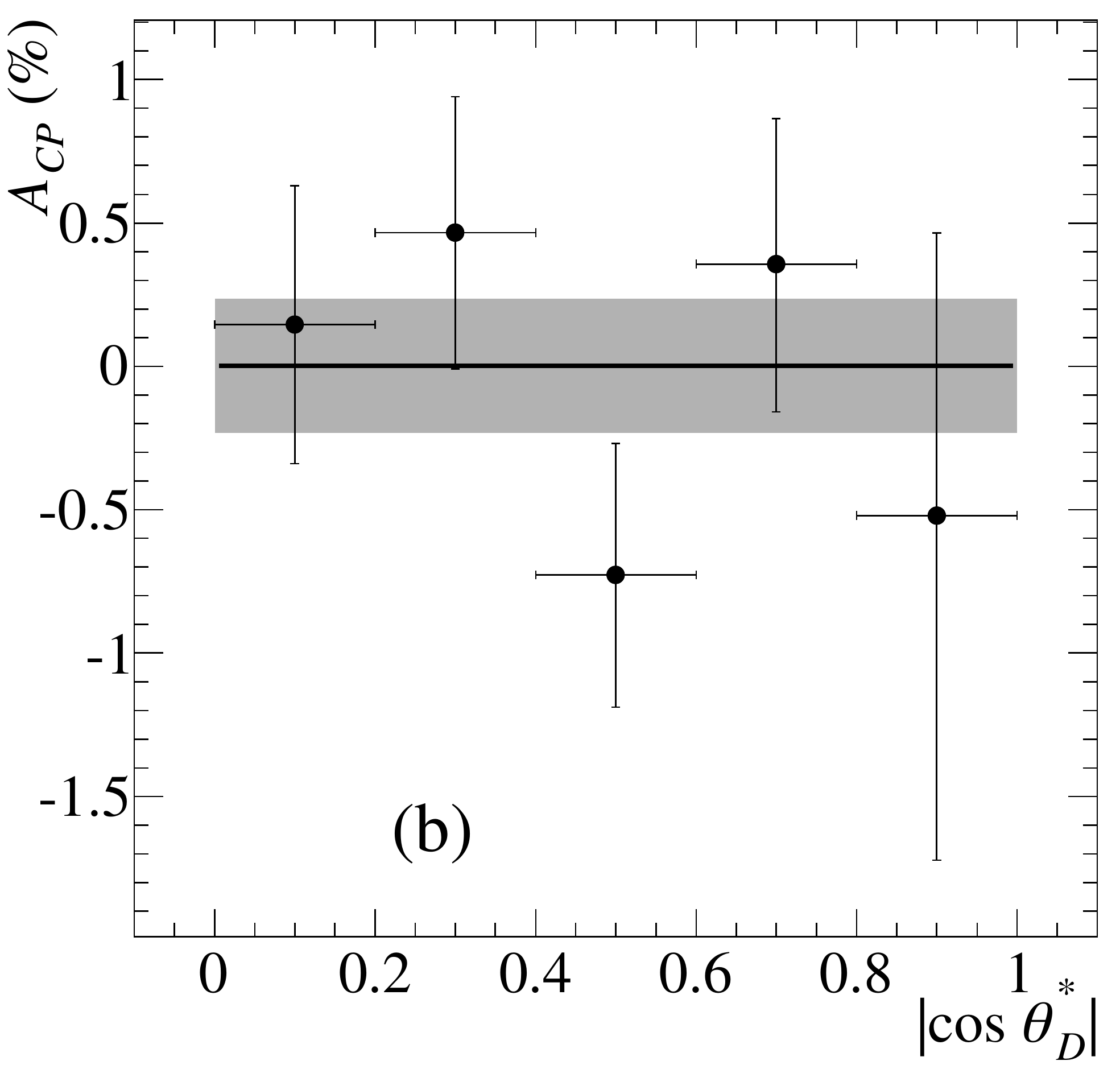} &
\includegraphics[width=0.33\textwidth,clip=true]{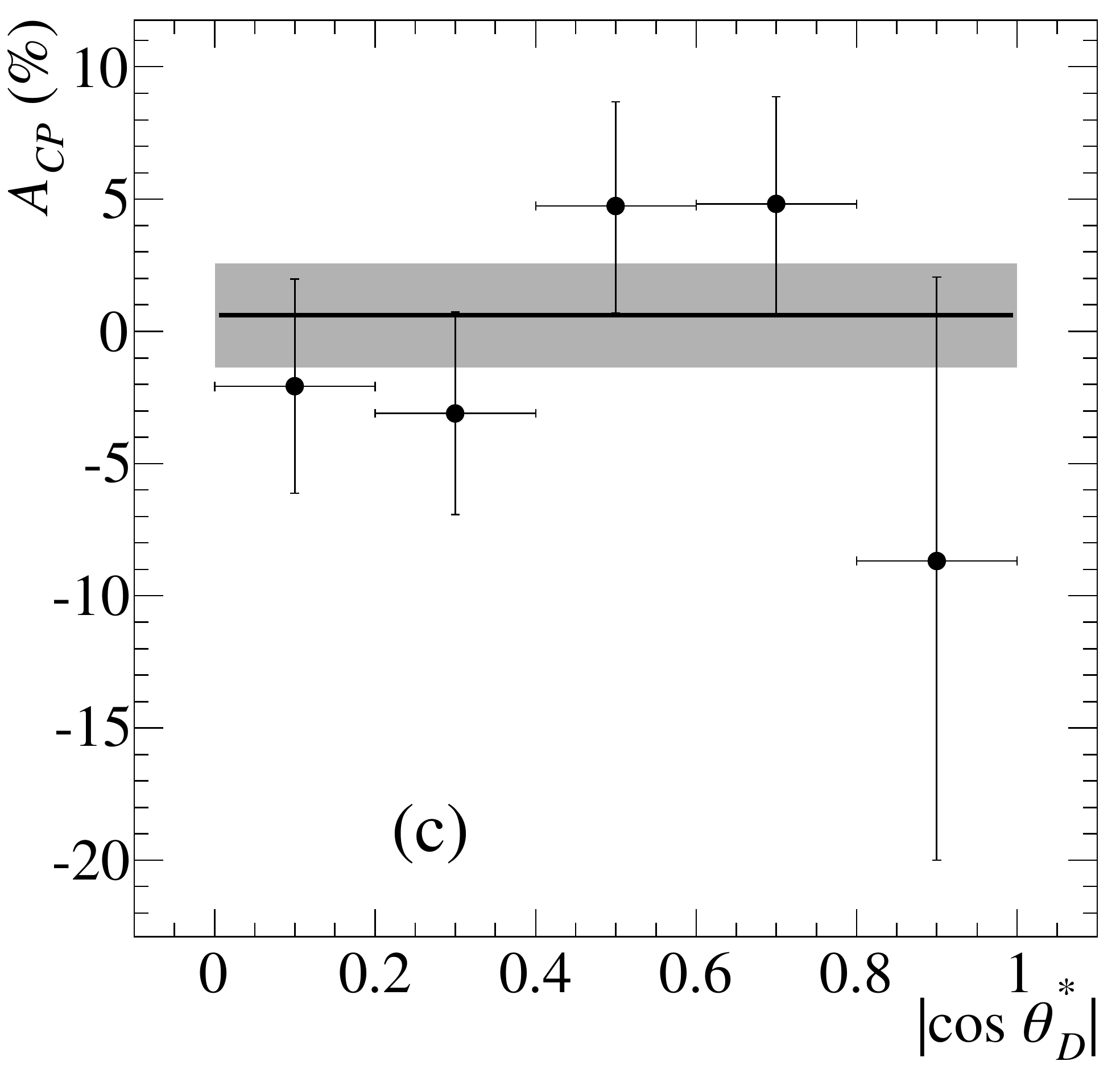} \\
\includegraphics[width=0.33\textwidth,clip=true]{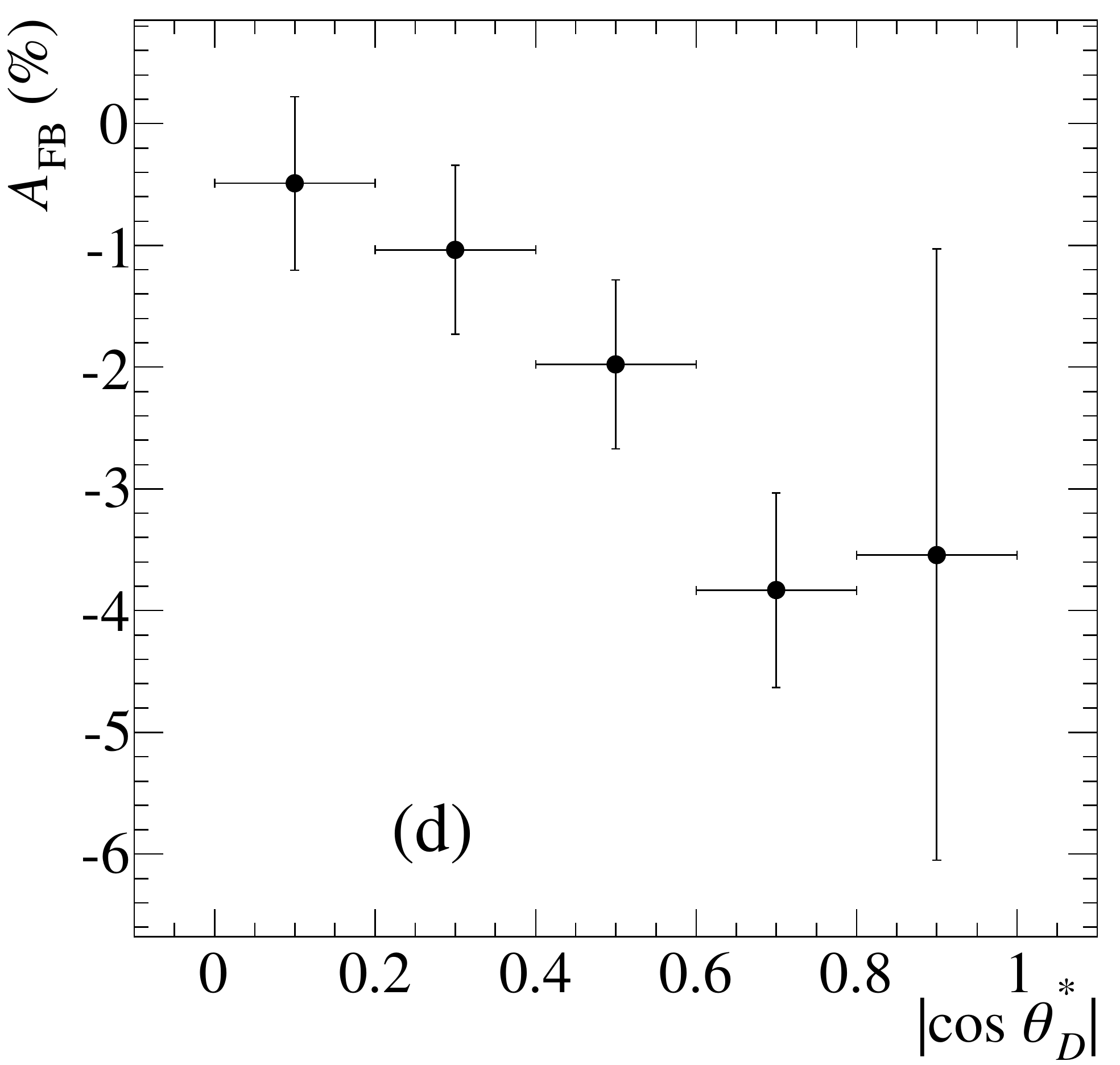} &
\includegraphics[width=0.33\textwidth,clip=true]{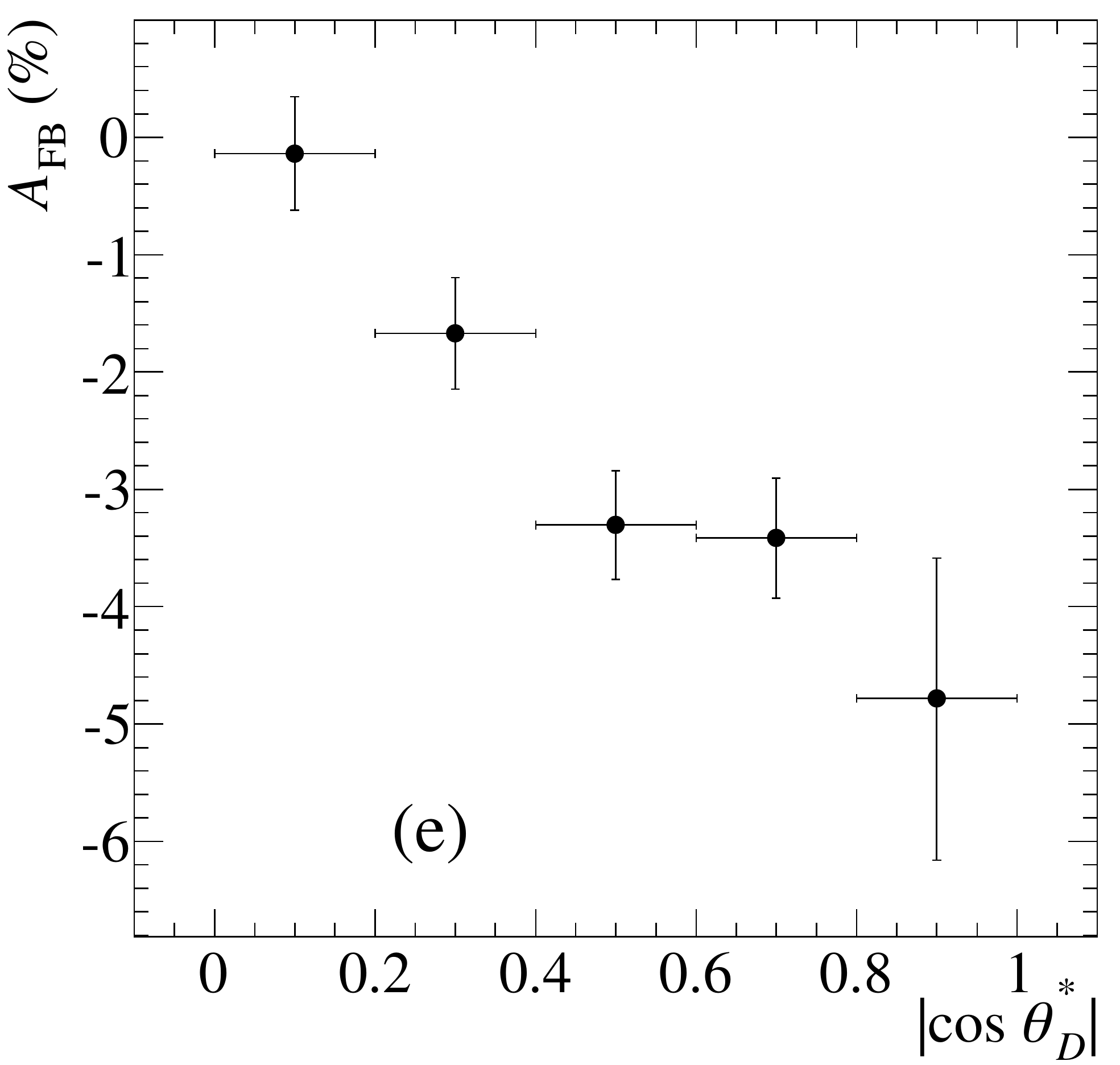} &
\includegraphics[width=0.33\textwidth,clip=true]{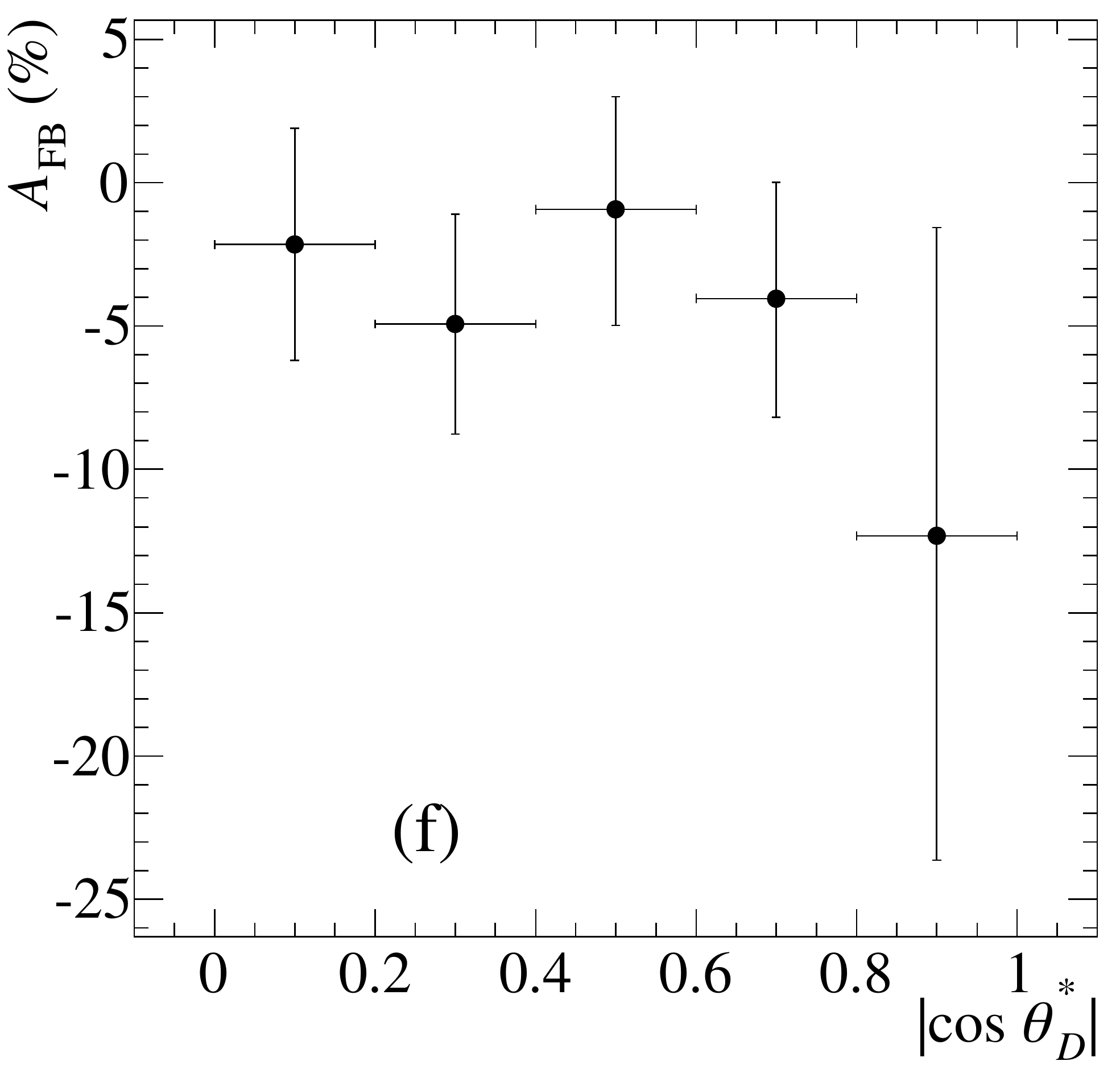} \\
\end{tabular}
\vspace{-0.3cm} \caption{
  \CP asymmetry, $A_{\CP}$, for (a) \Dtoksk, (b) \Dstoksk, and (c) \Dstokspi
  as a function of $|\cos\theta^*_D|$ in the data sample.
  The solid line represents the central value of $A_{\CP}$
  and the gray band is the $\pm1\,\sigma$ interval, both 
  obtained from a $\chi^2$-minimization
  assuming no dependence on $|\cos\theta^*_D|$.
  The corresponding forward-backward asymmetries, $A_{\FB}$, are shown in (d), (e), and (f).
}
\label{fig6} \vspace{-0.7cm} \end{center} \end{figure*}

The $A_{\CP}$ and $A_{\FB}$ values for the five $|\cos\theta^*_D|$ bins are shown 
in Fig.~\ref{fig6} for the three decay modes.
The weighted average of the five $A_{\CP}$ values is
$(0.16 \pm 0.36)\%$ for the \Dtoksk mode,
$(0.00 \pm 0.23)\%$ for \Dstoksk,
and $(0.6 \pm 2.0)\%$ for \Dstokspi,
where the errors are statistical only.

We perform two tests to validate the analysis procedure for each channel. 
The first involves generating 5000 toy MC experiments with a statistics equal to
data using the PDF and the parameters obtained from the fit to data.
After extracting $A_{\CP}$ from each experiment,
for the \Dtoksk and \Dstoksk modes, 
we deduce from the mean of the $A_{\CP}$ pull distributions the presence of a small bias
in the fitted value of each fit parameter 
(the means are $-0.036 \pm 0.014$ and $+0.041 \pm 0.014$, respectively). 
To account for this effect we apply a correction to the final values equal to
$+0.013\%$ for the \Dtoksk mode, and $-0.01\%$ for the \Dstoksk mode.
The $A_{\CP}$ pull distributions show that the fit provides an accurate estimate
of the statistical error for all the modes.
The second test involves fitting a large number of MC events from the 
full \babar detector simulation.
We measure $A_{\CP}$ from this MC sample to be consistent with the generated value of zero.

\section{Systematics}

The main sources of systematic uncertainty are listed in Table~\ref{tab_syst} 
for each decay mode, together with the overall uncertainties.
\begin{table*}[tb]
\caption{Summary of the systematic uncertainty contributions 
for the $A_{\CP}$ measurement in each mode.
The values are absolute uncertainties, even though given as percentages.
The total value corresponds to the sum in quadrature of the individual contributions.}
\begin{center}
\begin{tabular}{|l|c|c|c|}\hline
Systematic uncertainty & \Dtoksk & \Dstoksk & \Dstokspi \\
\hline
Efficiency of PID selectors          & 0.05\% & 0.05\% & 0.05\% \\
Statistics of the control sample  & 0.23\% & 0.23\% & 0.06\% \\
Misidentified tracks in the control sample    & 0.01\% & 0.01\% & 0.01\% \\
$\cos\theta^*_D$ interval size     & 0.04\% & 0.02\% & 0.27\% \\
\KzKzb regeneration                & 0.05\% & 0.05\% & 0.06\% \\
\KSKL interference                & 0.015\% & 0.014\% & 0.008\% \\
\hline
Total                                & 0.25\% & 0.24\% & 0.29\% \\
\hline
\end{tabular}
\label{tab_syst}
\end{center}
\end{table*}
The primary sources of systematic uncertainty are
the detection efficiency ratios used to weight the $\Dps^-$ yields, and 
the contributions from mis-identified particles in the data control sample used 
to determine the charge asymmetry in track reconstruction efficiency.

The technique used to remove the charge asymmetry due to detector-induced effects produces a 
small systematic uncertainty in the measurement of $A_{\CP}$ due to the statistical error in the 
relative efficiency estimation. This systematic uncertainty depends only on the type of 
charged particle (pion or kaon) in the final state, and not on the initial state. To 
estimate the systematic uncertainty on $A_{\CP}$ resulting from this source, the relative charged-particle 
efficiency in each interval of momentum and $\cos\theta$ is randomly drawn from a Gaussian 
distribution whose mean is the nominal relative efficiency in that interval, 
and where the root-mean-squared (r.m.s.) deviation is the 
corresponding statistical error. For each mode, we generate 500 such charged-particle 
relative-efficiency distributions, and use them to obtain 500 $A_{\CP}$ values, following the 
procedure described earlier to determine the nominal value of $A_{\CP}$. 
The r.m.s.~deviation of these 500 values from the nominal $A_{\CP}$ is taken to be the systematic uncertainty. For 
the \Dpstoksk modes, the estimated systematic uncertainty is 0.23\%. For the \Dstokspi
mode, we assign the same systematic uncertainty, 0.06\%, as that estimated for the
\Dtokspi mode in Ref.~\cite{delAmoSanchez:2011zza}.

The small fraction of misidentified particles in the generic track sample 
can introduce small biases in the estimation of the efficiencies,
and subsequently in the $A_{\CP}$ measurements.
Because of the good agreement between data and MC samples, we can use the simulated MC candidates 
to measure the shift in the $A_{\CP}$ value from the fit when the corrections are applied,
and when they are not.
Again, this contribution depends only on the type of the charged-particle track.
Hence, for the \Dstokspi mode, we assume the same shift obtained in Ref.~\cite{delAmoSanchez:2011zza}, namely +0.05\%. 
By fitting the \Dstoksk MC sample when the corrections are applied, and again when not,
we obtain a shift of +0.05\% and we assume this for both the \Dstoksk and \Dtoksk modes.
For all the modes, we shift the measured $A_{\CP}$ by this correction value
and then, conservatively, include the magnitude of this shift as a contribution to the systematic uncertainty.

Using MC simulation, we evaluate an additional systematic uncertainty
of $\pm0.01\%$ due to a possible charge asymmetry present in the control sample 
before applying the selection criteria.
Another source of systematic uncertainty is due to the choice of 
the $\cos\theta^*_D$ interval-size in the simultaneous ML fit. 
The systematic uncertainty is taken to be the largest absolute difference 
between the nominal $A_{\CP}$ extracted using ten $\cos\theta^*_D$ intervals 
and that obtained when the fit is performed using either 8 or 12 intervals in $\cos\theta^*_D$.
This is the dominant source of systematic uncertainty for the \Dstokspi mode, 
as shown in Table~\ref{tab_syst}.

We also consider a possible systematic uncertainty due to the regeneration
of neutral kaons in the material of the detector. 
The \Kz and \Kzb mesons produced in the decay processes can interact with the 
material in the tracking volume before they decay.
Following a method similar to that described in Ref.~\cite{Ko:2010mk},
we compute the probability for a \Kz or a \Kzb meson to interact 
inside the \babar tracking system, and
estimate systematic uncertainties of $0.05\%$ (\Dpstoksk) and $0.06\%$ (\Dstokspi).

\begin{table*}[tbp]
\caption{Summary of the $A_{\CP}$ measurements.
Where reported, the first uncertainty is statistical, and the second is systematic.} 
\begin{center}
\begin{tabular}{|l|c|c|c|}\hline
  & \Dtoksk & \Dstoksk & \Dstokspi \\
\hline
$A_{\CP}$ value from the fit      & $(+0.155 \pm 0.360)\%$ & $(0.00 \pm 0.23)\%$ & $(+0.6 \pm 2.0)\%$ \\ 
\hline
Correction for the bias from toy MC experiments & $+0.013\%$ & $-0.01\%$ & $-$ \\ 
Correction for the bias in the PID selectors & $-0.05\%$ & $-0.05\%$ & $-0.05\%$ \\
Correction for the \KSKL interference ($\Delta A_{\CP}$) & $+0.015\%$ & $+0.014\%$ & $-0.008\%$ \\
\hline
$A_{\CP}$ final value & $(+0.13 \pm 0.36 \pm 0.25)\%$ & $(-0.05 \pm 0.23 \pm 0.24)\%$ & $(+0.6 \pm 2.0 \pm 0.3)\%$\\
\hline
$A_{\CP}$ contribution from \KzKzb mixing
& $(-0.332 \pm 0.006)\%$ & $(-0.332 \pm 0.006)\%$ & $(+0.332 \pm 0.006)\%$ \\
\hline
$A_{\CP}$ final value (charm only) & $(+0.46 \pm 0.36 \pm 0.25)\%$ & $(+0.28 \pm 0.23 \pm 0.24)\%$ & $(+0.3 \pm 2.0 \pm 0.3)\%$\\
\hline
\end{tabular}
\label{tab_final}
\end{center}
\end{table*}

Although the intermediate state is labelled as a \KS, we apply a correction term to the 
measured $A_{\CP}$ to include the effect of \KSKL interference in the intermediate state~\cite{Grossman:2011zk}. 
This correction term depends on the proper time range over which decay distributions are integrated, and 
on the efficiency of the reconstruction of the $\pip\pim$ final state as a function of proper time. 
We compute the reconstruction efficiency distribution as a function of 
proper time using MC truth-matched \KS decays after the full selection. 
Following the method in Ref.~\cite{Grossman:2011zk} we estimate 
the asymmetry-correction term $\Delta A_{\CP}$ defined as:
\begin{equation}
\Delta A_{\CP} = A_{\CP}^{\textrm{corr}} - A_{\CP}^{\textrm{fit}},
\end{equation}
where $A_{\CP}^{\textrm{fit}}$ is the value obtained from the fit and
$A_{\CP}^{\textrm{corr}}$ is the corrected value.
The correction terms are reported in Table~\ref{tab_final}
and, to be conservative, we include their absolute values
as contributions to the systematic uncertainty estimates.
We also estimate the correction factor for the \Dtokspi mode 
using the \KS reconstruction efficiency distribution after the selection detailed 
in Ref.~\cite{delAmoSanchez:2011zza}, and obtain the value +0.002\%.
All these corrections are rather small, even compared to those estimated in a
similar analysis~\cite{BABAR:2011aa}. 
The smaller values of the corrections in the present analysis are due to the 
improved efficiency for \KS mesons with short decay times that we obtain 
by applying the requirement on the decay length divided by its uncertainty, 
rather than on the decay length alone.

\section{Conclusion}

In conclusion, we measure the direct \CP asymmetry, $A_{\CP}$, in the 
\Dtoksk, \Dstoksk, and \Dstokspi modes using approximately 
159\,000, 288\,000, and 14\,000 
signal candidates, respectively.
The measured $A_{\CP}$ value for each mode is reported in Table~\ref{tab_final},
where the first errors are statistical and the second are systematic.  
In the last row of the table, we also report  the $A_{\CP}$ values after 
subtracting the expected $A_{\CP}$ contribution for each mode due to \KzKzb mixing.
The results are consistent with zero, and with the SM prediction,
within one standard deviation.

\section{Acknowledgements}

We are grateful for the 
extraordinary contributions of our \pep2 colleagues in
achieving the excellent luminosity and machine conditions
that have made this work possible.
The success of this project also relies critically on the 
expertise and dedication of the computing organizations that 
support \babar.
The collaborating institutions wish to thank 
SLAC for its support and the kind hospitality extended to them. 
This work is supported by the
US Department of Energy
and National Science Foundation, the
Natural Sciences and Engineering Research Council (Canada),
the Commissariat \`a l'Energie Atomique and
Institut National de Physique Nucl\'eaire et de Physique des Particules
(France), the
Bundesministerium f\"ur Bildung und Forschung and
Deutsche Forschungsgemeinschaft
(Germany), the
Istituto Nazionale di Fisica Nucleare (Italy),
the Foundation for Fundamental Research on Matter (The Netherlands),
the Research Council of Norway, the
Ministry of Education and Science of the Russian Federation, 
Ministerio de Ciencia e Innovaci\'on (Spain), and the
Science and Technology Facilities Council (United Kingdom).
Individuals have received support from 
the Marie-Curie IEF program (European Union) and the A. P. Sloan Foundation (USA).

\end{document}

%% file: authors_oct2012_bad2480.tex
%
\author{J.~P.~Lees}
\author{V.~Poireau}
\author{V.~Tisserand}
\affiliation{Laboratoire d'Annecy-le-Vieux de Physique des Particules (LAPP), Universit\'e de Savoie, CNRS/IN2P3,  F-74941 Annecy-Le-Vieux, France}
\author{E.~Grauges}
\affiliation{Universitat de Barcelona, Facultat de Fisica, Departament ECM, E-08028 Barcelona, Spain }
\author{A.~Palano$^{ab}$ }
\affiliation{INFN Sezione di Bari$^{a}$; Dipartimento di Fisica, Universit\`a di Bari$^{b}$, I-70126 Bari, Italy }
\author{G.~Eigen}
\author{B.~Stugu}
\affiliation{University of Bergen, Institute of Physics, N-5007 Bergen, Norway }
\author{D.~N.~Brown}
\author{L.~T.~Kerth}
\author{Yu.~G.~Kolomensky}
\author{G.~Lynch}
\affiliation{Lawrence Berkeley National Laboratory and University of California, Berkeley, California 94720, USA }
% ...
\collaboration{The \babar\ Collaboration}
\noaffiliation